\documentclass[11pt,a4paper]{article}
\usepackage[utf8]{inputenc}
\usepackage{amsmath}
\usepackage{amsfonts}
\usepackage{amssymb}
\usepackage{bm}
\usepackage{pdflscape}
\usepackage[margin=1in,footskip=0.25in]{geometry}
\usepackage{graphicx}
\usepackage{multirow}
\usepackage{graphicx}
\usepackage{lscape}
\usepackage[table,xcdraw]{xcolor}
\usepackage{float}

\usepackage{longtable}
\begin{document}
\noindent\textbf{\Large{Selection of link function in binary regression: A case-study with world happiness report on immigration}}
\\
\\
Ardhendu Banerjee\textsuperscript{a}\,\,\ ,\,\,Subrata Chakraborty\textsuperscript{b} \,\,and\,\, Aniket Biswas\textsuperscript{c}\\
\textsuperscript{a}Department of Statistics, Calcutta University, Kolkata.\\
\textsuperscript{b,c}Department of Statistics, Dibrugarh University, Dibrugarh.
\\
Email:\,\,\textsuperscript{a}\textit{rana.ardh1024@gmail.com} \,\,\,\,\,\textsuperscript{b}\textit{subrata\_stats@dibru.ac.in}
\,\,\,\,\,\textsuperscript{c}\textit{biswasaniket44@gmail.com}
\section*{Abstract}
Selection of appropriate link function for binary regression remains an important issue for data analysis and its influence on related inference. We prescribe a new data-driven methodology to search for the same, considering some popular classification assessment metrics. A case-study with World Happiness report,2018 with special reference to immigration is presented for demonstrating utility of the prescribed routine. \\
\\
\noindent \textbf{Keywords}: Binary regression, Link function, Latent stress-strength, World happiness report, Cross-validation.\\
\\
\noindent \textbf{2010 MSC}: 62-07. 62J12. 62P25.
\section{Introduction}
\begin{center}
    ``Happiness is the joy that we feel when we’re striving after our potential."\\
    - Ancient Greek World
    \end{center} 
    
 \noindent In 1979, at Bombay Airport, the King of Bhutan, Jigme Singye Wangchuck, replied to a query from one Indian journalist that, `` We do not believe in Gross National Product because Gross National Happiness is more important ". This was the beginning of the great philosophy as reported by  Tashi Dorji, in his famous article,`` The story of a king, a poor country and a rich idea "- in 2012 (see Dorji, 2012). UN started to follow this philosophy and as a result the Sixth World Happiness Report of 2018 since 2012 on the basis of Gallup World Poll(GWP) Data 2005-2017, being published. World Happiness Index measures the happiness level of the 156 countries and ranks them accordingly. Among these 156 countries, this report includes the happiness ranking of 117 countries on the basis of immigration (see Helliwell, 2018). This report considered a Cantrill Ladder (an imaginary ladder with 0 to 10 steps from bottom to top indicating the increasing level of happiness with higher steps) to measure the level of happiness. Parameters viz. Life Ladder, Log GDP per capita, Social support, Healthy life expectancy at birth, Freedom to make life choices, Generosity, Perceptions of corruption, Positive affect(the average of previous-day affect measures for happiness, laughter, and enjoyment), Negative affect (the average
of previous-day affect measures for worry, sadness, and anger), Confidence in national government, Democratic Quality and Delivery Quality, for better understanding of well-being are also reported country-wise. \\
\\
Nowadays glocal world is experiencing the wave of heavy migration. People have migrated to different countries due to different reasons according to their perceived level of aspiration. The primary question, while deciding to migrate to a new country with new environment, society, culture, habits and unknown people surrounding, what will be the best choice for destination? How one can decide the place of destination, which suits him/her best? \\

From statistical perspective of this study, happiness score is the concerned response variable and the various parameters mentioned above are covariates. Chapter 2 of World Happiness Report(2018) focuses on international migration. In statistical appendices of this chapter, pooled ordinary least square regression is performed for assessing the impact of each covariate on the response. It is to be noted that, the response variable here is score on an ordinal scale and thus it does not comply with the usual assumptions of ordinary least squares. A more technically correct thing would be to categorize each country into two categories: happy or unhappy based on the score and perform binary regression for the same purpose as reported in the mentioned statistical appendices while this will result in marginal loss of information in response variable. This being said, categorization of ordinal response is one way-out for drawing statistically correct inference. Question remains: why two categories? Choice of number of categories is not rigid but while doing this, natural intuition leads one to the dichotomy of happiness and unhappiness. Extension to more than two categories may make the physical interpretations clumsy. For performing binary regression, many link functions are available in literature both symmetric (e.g probit link) and asymmetric (e.g complementary log-log) but only some are popular. For details on binary regression and use of link functions, see Cox(2018) and Agresti and Kateri(2011). The choice of proper link function is important as mis-specification of the same might have adverse effect on inference and prediction. Some notable attempts for making a statistical choice for the appropriate link function can be found in Czado and Santner(1992), Huettmann and Linke(2003) and Jingwei(2014). In the current work we consider cross-validation based approach along with a number of important assessment indices to get a data-dependent choice of link function. Cross-validation based approaches are taken up in this work in view of the prediction purpose of binary regression modelling. We analyze multiple data-sets from the same context to find out the appropriate choice of link function and refrain from reporting the significance of individual covariates which was easily done once the suitable link was established.             We hope from the best fitted model we can improve the decision of choosing the country and increase the level of confidence of the immigrant that will be effective for living a good life with increasing potentiality for mankind. Motivation behind working with this data-set:
\begin{itemize}
\item Reliable, collected by a proven survey-group.
\item With moderate to large number of observations, cross-validation and thus the prescribed routine performs well. Thus, demonstration of the routine by considering this data-set is valid.
\item ``Happiness", in its true sense is a determinant of immigration.
\item Happiness index is the most talked after  indicator in recent times.  
\end{itemize} 

In the next section we formulate the statistical problem associated with the data-set. In section 3, we briefly discuss different assessment metrics used in this paper. Section 4 discusses two important cross-validation schemes. Then we present a note on the concerned survey and nature of covariates present in the study. In section 6, construction of working data-sets and findings from numerical results are given. We finish with a short discussion on relevance and scope of the study. Relevant tables are are provided in appendix.
\section{Formulation of the problem} 
Suppose, data on $n$ different countries where, the response variable ($S_i$) for $i$-th country  along with $k$ covariates ($\bm{z}_i$) are given for $i=1, 2, ..., n$. As mentioned in section 1, the response $S$ is ordinal and it is a well known fact that, such responses cannot be modelled efficiently with usual linear and generalized linear models (see Crichtonn and Hinde, 1992). Thus, with some fixed threshold $t$ we consider, 
\begin{equation*}
    U=\begin{cases}
    1\quad \textrm{if} \quad s\geq t\\
    0\quad \textrm{if} \quad s<t
    \end{cases}
\end{equation*}
which is a reflection of two contradictory latent forces viz. $X$ and $Y$ as follows:
\begin{equation*}
    P(U=1)=P(Y<X)
\end{equation*}
If we take $U=1$ as success, a natural interpretation for $X$ would be the positive force and $Y$ would be the negative force. Latency of $X$ and $Y$ prevents one from directly modelling $(X,Y)$. The set of covariates is naturally partitioned into $\bm{z}_1$ and $\bm{z}_2$, the former influencing $X$ and the latter $Y$.\\

Our interest here is to explore and investigate the possibilities of modeling of unobserved $X$ and $Y$. One can assume both to follow independent normal, logistic, cauchy or extreme value distribution among others. The problem of choosing appropriate model is equivalent to that of choosing link function for binary regression where, $P(U_i=1)$ is modelled as:
\begin{equation*}
P(U_i=1)=F^{-1}(\bm{z}_i^\prime\bm{\beta})
\end{equation*} 
where, $\bm{\beta}$ is the vector of parameters and $F^{-1}$ is a link function induced by the assumed common latent stress-strength distribution $F$. In this study, we consider the following four link functions as:
\begin{itemize}
\item Probit link: $F$ is cdf of standard normal distribution.
\item Logit link: $F$ is cdf of standard logistic distribution.
\item Cauchit link: $F$ is cdf of standard Cauchy distribution.
\item Complementary log-log link: $F$ is cdf of standard extreme value distribution.
\end{itemize} 
For observed $\{(U_i, \bm{z}_i): i=1, 2, ..., n\}$, the likelihood function involves chosen structure of $F^{-1}$ and maximum likelihood estimator for $\bm{\beta}$ is obtained through iterative re-weighted least squares. 
For details on connection of latent variable modelling with link functions see Cox(2018), Albert and Chib(1993) and Banerjee and Biswas(2003). Here we will apply some known methods for assessing suitability of link functions and simultaneously use cross validation to achieve desired level of predictive performance for the models. 

\section{Different measures of assessment}
In order to assess performance of link functions tried in our case study, we shall employ well known assessment measures available in literature (see Tharwart, 2018).
In binary classification problems, prediction of one of the two classes (usually +ve and -ve) is based on a new set of covariates. A positive (negative) sample point classified as positive (negative) is referred as True Positive(negative) classification whereas a positive (negative) sample point classified as negative (positive) is called false negative(positive) or Type II Error. The corresponding confusion matrix is shown below:\\

\begin{table}[H]
\center
\begin{tabular}{cccc}
\multicolumn{4}{c}{\hspace{4.9cm}\textbf{True or Actual Class}}\\
\\
&&Positive&Negative\\
\\
\multirow{2}{*}{\textbf{Predicted Class}}&True&True Positive (TP)& False Positive (FP)\\
\\
&False&False Negative (FN)& True Negative (TN)\\

\end{tabular}
\end{table}

\noindent Based on the confusion matrix, we consider the following four metrics:
\begin{eqnarray*}
   E_1&:& \textrm{Accuracy rate} = \frac{\textrm{no. of correct classification}}{\textrm{sample size}} = \frac{TP + TN}{P + N}\\
    E_2&:& \textrm{Sensitivity}= \frac{\textrm{no. of true positive classification}}{\textrm{total no. of positive classification}} = \frac{TP}{P}\\
    E_3&:& \textrm{Specificity}= \frac{\textrm{no. of true negative classification }}{\textrm{total no. of negative  classification}} = \frac{TN}{N}\\
 \end{eqnarray*}

\noindent $E_1$ is the simplest and commonly used measure and it is sensitive to imbalanced data. On the contrary $E_2$ and $E_3$ are not sensitive to imbalanced data. When mis-classifying true positive is more serious error than mis-classifying true negative, we should decide upon $E_2$ and for the opposite scenario, $E_3$ should be the metric to note. Along with these three, we also consider another popular assessment metric:
\begin{equation*}
    E_4 : \textrm{Area under the ROC curve (AUC)}
\end{equation*}
This measure is based on receiver operating characteristics (ROC) curve and overcomes the inability of the ROC in comparing different classifiers for being a scalar rather than a function itself. 
\section{On cross-validation methods}
 Since main reason for modelling here, is to predict probability of being happy, it is inevitable to subject the proposed models to rigorous cross validation for achieving perfection, in addition to the assessments discussed in last section. In this study, we implement three useful cross-validation routines, briefly discussed below. For details on various cross-validation approaches and their relative performance, see section 5.1 of James et al.(2013).\\
\begin{itemize}
    \item \textbf{Leave-p-out CV}:
This Method comprises of using $p$ out of $n$ observations as the validation set while the rest  $(n-p)$ observations are taken as training set. This exercise is repeated in all possible ways to partition a sample of $n$ into two sets, one with $p$ and the other with $n-p$ elements. Obviously with large $n$ and even moderate $p$, the number of validation sets $n\choose p$ may explode with $n$. For $p=1$, this method reduces to Leave-one-out cross validation (LOOCV). In numerical study, we apply LOOCV and LPOCV with two different choices of $p$: Hold $75\%$-Leave $25\%$ and Hold $50\%$-Leave $50\%$. 

\item \textbf{k-fold CV}: One way to avoid exhaustive CV method as above is to apply k- fold CV, where the sample is randomly partitioned into k sub-samples of same size. Out of these k sub-samples, one is taken as the validation set and rest are used for training. This process is repeated such that each of the k sub-sample are taken as training set. In numerical study, we perform this with $k=5$ and $k=10$.
\end{itemize}
    
\section{A note on GWP survey}
Gallup World Poll (GWP) conducted surveys over 160 countries since 2005 with 1000 sample (for large countries it can be of size 2000) of adult population semiannually, annually, and biennially . This survey includes almost 100 questions in a similar manner for the people of different region of world either through telephone (generally in the developed countries) for almost 30 minutes or direct interview (generally in developing countries) for almost 1 hour. World Happiness Report (WHR), 2018 used this GWP data for developing the happiness index, and modelling it with the  covariates  given and described in Table 5. In accordance with the formulation given in section 2, we identify the factors which contribute positively towards happiness of migrants to be covariates for $X$ and the remaining as covariates for $Y$ and the same is given in column 3 of Table 5.  \\    
\section{Numerical study and findings}
We consider four different but related data-sets for demonstrating the method of choosing the suitable link function or equivalently, latent stress-strength models. As mentioned in section 2, data-sets reporting happiness score with related covariates are available for different years. Here, we consider these data-sets for 2015, 2016 and 2017. For stability, we synthetically reproduce another data-set with the current year happiness score along with the covariate-values averaged over the years 2005 to 2017. The 2017 data-set contains information on 10 covariates (excluding positive effect and negative effect) while the other data-sets contain all 12 covariates. The working data-sets (https://worldhappiness.report/ed/2018/) have been used for ranking the countries according to the migrants' satisfactory level using the happiness score. As indicated in section 1 and section 2, we categorize the countries to be a good choice for migration from ``happiness" perspective if the score is greater than or equal to 6. Remaining countries fall into the other category. Thus, the transformed binary response variable, $U$ is as follows:
\begin{equation*}
    U=\begin{cases}
    1\quad\textrm{if} \quad\textrm{happiness score}\geq 6\\
    0\quad\textrm{if} \quad\textrm{happiness score}< 6
    \end{cases}
\end{equation*}
Obviously, the transformed variable $U$ is response in the current study. Our main interest is to model $P(U=1)$ with available covariates using different link functions and to find out the most suitable one.\\

For each of the four data-sets discussed above, we compute different assessment metrics $E_1$, $E_2$, $E_3$ and $E_4$ with $5$ cross-validation routine mentioned in section 4. The numerical results for years 2017, 2016, 2015 and the aggregate data are given in Table 1, Table 2, Table 3 and Table 4, respectively. For LOOCV, the metrics except $E_1$ cannot be calculated as, the single test sample will either be $1$ or $0$.\\

With respect to each metric of assessment, we identify the best-performing link function as the one which has the maximum number of rank $1$ over all $5$ cross-validation routines. Tie if any, is resolved by going to the next stage and checking for the next rank and so on. Using this scheme, we arrive at the following conclusions:\\

\noindent With respect to $E_1$:
\begin{itemize}
    \item For 2017, logit performs best followed by probit. 
    \item For 2016, cauchit performs best followed by complementary log-log.
    \item For 2015, cauchit performs best followed by complementary log-log.
    \item For aggregate data, complementary log-log performs best followed by probit.
\end{itemize}

\noindent With respect to $E_2$:
\begin{itemize}
    \item For 2017, probit performs best followed by logit. 
    \item For 2016, cauchit performs best followed by logit and probit.
    \item For 2015, cauchit performs best followed by logit.
    \item For aggregate data, logit performs best followed by probit.
\end{itemize}

\noindent With respect to $E_3$:
\begin{itemize}
    \item For 2017, complementary log-log performs best followed by cauchit. 
    \item For 2016, cauchit and complementary log-log both performs equally well.
    \item For 2015, complementary log-log performs best followed by cauchit.
    \item For aggregate data, cauchit performs best followed by probit.
\end{itemize}

\noindent With respect to $E_4$:
\begin{itemize}
    \item For 2017, cauchit performs best followed by logit. 
    \item For 2016, probit performs best followed by logit.
    \item For 2015, cauchit and complementary log-log performs equally well.
    \item For aggregate data, probit performs best followed by logit.
\end{itemize}

Overall it is observed that the cauchit link function is the best or second best in half of the cases followed by complementary log-log link. Practitioners of binary regression modelling should therefore give efforts to search for the best one from a set of available link functions for better inference and prediction.
 \section{Discussion}
 It is true that, there is a tendency among analysts to opt for logit link function while dealing with binary response modelling despite the fact that the distributional assumptions underlying such choice of link function may not hold very often. This has a potential of generating statistically incorrect findings and consequences may be costly in some domain of research. The findings of our investigation confirms the issue and highlights how different link functions come upfront surpassing the established myths with data from the same context and with respect to different periods and assessment metric. The data driven methodology to look for the best link function presented in this short case study aims to provide a meaningful way to address the issue. 
 \section*{References}
 Agresti, A., \& Kateri, M. (2011). \textit{Categorical data analysis}. Springer Berlin Heidelberg.\\
\\
Albert, J. H., \& Chib, S. (1993). Bayesian analysis of binary and polychotomous response data. \textit{Journal of the American statistical Association}, 88(422), 669-679.\\
\\
Banerjee, T., \& Biswas, A. (2003). A new formulation of stress–strength reliability in a regression setup. \textit{Journal of statistical planning and inference}, 112(1-2), 147-157.\\
\\
Cox, D. R. (2018). \textit{Analysis of binary data}. Routledge.\\
\\
Crichton, N., \& Hinde, J. (1992). Investigation of an ordered logistic model for consumer debt. In Advances in GLIM and \textit{Statistical Modelling} (pp. 54-59). Springer, New York, NY.\\
\\
Czado, C., \& Santner, T. J. (1992). The effect of link misspecification on binary regression inference. \textit{Journal of statistical planning and inference}, 33(2), 213-231.\\
\\
Dorji, T. (2012). \textit{https://blogbhutan.wordpress.com/2012/06/11/the-story-of-a-king-a-poor-country-and-a-rich-idea/}\\
\\
Helliwell, J., Layard, R., \& Sachs, J. (2018). World Happiness Report 2018, \textit{New York: Sustainable Development Solutions Network}\\
\\
Huettmann, F., \& Linke, J. (2003, May). Assessment of different link functions for modeling binary data to derive sound inferences and predictions. In \textit{International Conference on Computational Science and Its Applications} (pp. 43-48). Springer, Berlin, Heidelberg.\\
\\
James, G., Witten, D., Hastie, T., \& Tibshirani, R. (2013). \textit{An introduction to statistical learning} (Vol. 112, p. 18). New York: Springer.\\
\\
Li, J. (2014). \textit{Choosing the proper link function for binary data} (Doctoral dissertation). Link: \textit{https://repositories.lib.utexas.edu/handle/2152/26363}\\
\\
Tharwat, A. (2018). Classification assessment methods. \textit{Applied Computing and Informatics}.\\
\\
\newpage
\section*{Appendix}
\begin{longtable}{ccccc}
\caption{Based on 2017 data-set}\\
\endfirsthead

\hline\\
\multicolumn{5}{c}{LOOCV}\\\\
\hline
Efficiency Measure &Probit&Logit&Cauchit&C-Log-Log\\
\hline
$E_1$&0.92437&0.92437&0.91597&0.91597\\
$E_2$&-&-&-&-\\
$E_3$&-&-&-&-\\
$E_4$&-&-&-&-\\
\hline \\
\multicolumn{5}{c}{LPOCV 50-50 }\\\\
\hline
Efficiency Measure &Probit&Logit&Cauchit&C-Log-Log\\
\hline
$E_1$&0.87237&0.87187&0.87790&0.87373\\
$E_2$&0.84482&0.84403&0.83985&0.79121\\
$E_3$&0.88116&0.88157&0.89471&0.91237\\
$E_4$&0.86316&0.86325&0.86555&0.84927\\
\hline\\
\multicolumn{5}{c}{LPOCV 75-25}\\\\
\hline
Efficiency Measure &Probit&Logit&Cauchit&C-Log-Log\\
\hline
$E_1$&0.89867&0.89930&0.89880&0.89550\\
$E_2$&0.87655&0.87454&0.85686&0.81494\\
$E_3$&0.91414&0.91528&0.91872&0.93653\\
$E_4$&0.89235&0.89235&0.88829&0.87556\\
\hline\\
\multicolumn{5}{c}{5 Fold CV}\\\\
\hline
Efficiency Measure &Probit&Logit&Cauchit&C-Log-Log\\
\hline
$E_1$&0.92437&0.88235&0.90756&0.89916\\
$E_2$&0.87507&0.90728&0.89931&0.83459\\
$E_3$&0.88805&0.90810&0.92839&0.95928\\
$E_4$&0.92180&0.90560&0.88214&0.91859\\
%\end{tabular}
\hline\\
\multicolumn{5}{c}{10 Fold CV}\\\\
\hline
Efficiency Measure &Probit&Logit&Cauchit&C-Log-Log\\
\hline
$E_1$&0.88225&0.92437&0.89076&0.91597\\
$E_2$&0.87395&0.74789&0.83673&0.92941\\
$E_3$&0.91176&0.93389&0.92577&0.92913\\
$E_4$&0.88229&0.87057&0.89675&0.89499\\
\hline
\end{longtable}

\begin{longtable}{ccccc}
\caption{Based on 2016 data-set}\\
\endfirsthead

\hline\\
\multicolumn{5}{c}{LOOCV}\\\\
\hline
Efficiency Measure &Probit&Logit&Cauchit&C-Log-Log\\
\hline
$E_1$&0.88618&0.88618&0.91057&0.89431\\
$E_2$&-&-&-&-\\
$E_3$&-&-&-&-\\
$E_4$&-&-&-&-\\
\hline \\
\multicolumn{5}{c}{LPOCV 50-50 }\\\\
\hline
Efficiency Measure &Probit&Logit&Cauchit&C-Log-Log\\
\hline
$E_1$&0.85126&0.85197&0.86955&0.85765\\
$E_2$&0.81341&0.81373&0.81845&0.75867\\
$E_3$&0.87120&0.87292&0.89362&0.90662\\
$E_4$&0.84398&0.84503&0.85686&0.83296\\
\hline\\
\multicolumn{5}{c}{LPOCV 75-25}\\\\
\hline
Efficiency Measure &Probit&Logit&Cauchit&C-Log-Log\\
\hline
$E_1$&0.87861&0.84800&0.88290&0.88119\\
$E_2$&0.84960&0.84794&0.84388&0.82313\\
$E_3$&0.89580&0.89557&0.90249&0.91086\\
$E_4$&0.87036&0.86985&0.86929&0.86545\\
\hline\\
\multicolumn{5}{c}{5 Fold CV}\\\\
\hline
Efficiency Measure &Probit&Logit&Cauchit&C-Log-Log\\
\hline
$E_1$&0.88618&0.89431&0.90244&0.90244\\
$E_2$&0.87434&0.82488&0.90129&0.77108\\
$E_3$&0.91578&0.89938&0.94026&0.92898\\
$E_4$&0.85664&0.86206&0.87976&0.86569\\
%\end{tabular}
\hline\\
\multicolumn{5}{c}{10 Fold CV}\\\\
\hline
Efficiency Measure &Probit&Logit&Cauchit&C-Log-Log\\
\hline
$E_1$&0.89434&0.90244&0.90244&0.92683\\
$E_2$&0.89282&0.89377&0.85528&0.85163\\
$E_3$&0.89649&0.89959&0.90646&0.92203\\
$E_4$&0.88272&0.84327&0.86956&0.87964\\
\hline
\end{longtable}

\begin{longtable}{ccccc}
\caption{Based on 2015 data-set}\\
\endfirsthead

\hline\\
\multicolumn{5}{c}{LOOCV}\\\\
\hline
Efficiency Measure &Probit&Logit&Cauchit&C-Log-Log\\
\hline
$E_1$&0.91869&0.91057&0.91057&0.91867\\
$E_2$&-&-&-&-\\
$E_3$&-&-&-&-\\
$E_4$&-&-&-&-\\
\hline \\
\multicolumn{5}{c}{LPOCV 50-50 }\\\\
\hline
Efficiency Measure &Probit&Logit&Cauchit&C-Log-Log\\
\hline
$E_1$&0.87959&0.88045&0.90095&0.87831\\
$E_2$&0.85221&0.85256&0.86183&0.75874\\
$E_3$&0.89429&0.89517&0.92173&0.93732\\
$E_4$&0.87642&0.87669&0.89002&0.85069\\
\hline\\
\multicolumn{5}{c}{LPOCV 75-25}\\\\
\hline
Efficiency Measure &Probit&Logit&Cauchit&C-Log-Log\\
\hline
$E_1$&0.89487&0.89554&0.90303&0.89390\\
$E_2$&0.87261&0.87254&0.87889&0.83568\\
$E_3$&0.91338&0.91342&0.92243&0.93295\\
$E_4$&0.88657&0.88759&0.89545&0.87789\\
\hline\\
\multicolumn{5}{c}{5 Fold CV}\\\\
\hline
Efficiency Measure &Probit&Logit&Cauchit&C-Log-Log\\
\hline
$E_1$&0.87801&0.86992&0.91869&0.94309\\
$E_2$&0.86267&0.91722&0.91223&0.81844\\
$E_3$&0.91540&0.91054&0.88686&0.96386\\
$E_4$&0.86579&0.88979&0.84357&0.89257\\
%\end{tabular}
\hline\\
\multicolumn{5}{c}{10 Fold CV}\\\\
\hline
Efficiency Measure &Probit&Logit&Cauchit&C-Log-Log\\
\hline
$E_1$&0.90244&0.90244&0.90244&0.92683\\
$E_2$&0.88880&0.80434&0.888577&0.92547\\
$E_3$&0.91426&0.90871&0.90089&0.93413\\
$E_4$&0.89381&0.90527&0.89636&0.92222\\
\hline
\end{longtable}

\begin{longtable}{ccccc}
\caption{Based on the aggregate data-set}\\
\endfirsthead

\hline\\
\multicolumn{5}{c}{LOOCV}\\\\
\hline
Efficiency Measure &Probit&Logit&Cauchit&C-Log-Log\\
\hline
$E_1$&0.90647&0.89928&0.88489&0.89928\\
$E_2$&-&-&-&-\\
$E_3$&-&-&-&-\\
$E_4$&-&-&-&-\\
\hline \\
\multicolumn{5}{c}{LPOCV 50-50 }\\\\
\hline
Efficiency Measure &Probit&Logit&Cauchit&C-Log-Log\\
\hline
$E_1$&0.85891&0.85887&0.86636&0.86317\\
$E_2$&0.81299&0.81192&0.80615&0.76611\\
$E_3$&0.88721&0.88719&0.89612&0.91165\\
$E_4$&0.84916&0.84902&0.85106&0.83805\\
\hline\\
\multicolumn{5}{c}{LPOCV 75-25}\\\\
\hline
Efficiency Measure &Probit&Logit&Cauchit&C-Log-Log\\
\hline
$E_1$&0.88414&0.88363&0.87677&0.88368\\
$E_2$&0.82103&0.81873&0.80982&0.77989\\
$E_3$&0.91621&0.91539&0.90996&0.93042\\
$E_4$&0.87367&0.87269&0.86159&0.85784\\
\hline\\
\multicolumn{5}{c}{5 Fold CV}\\\\
\hline
Efficiency Measure &Probit&Logit&Cauchit&C-Log-Log\\
\hline
$E_1$&0.87769&0.83453&0.89928&0.90647\\
$E_2$&0.77307&0.85172&0.81814&0.76321\\
$E_3$&0.88082&0.90521&0.90692&0.92817\\
$E_4$&0.89934&0.83118&0.85167&0.87317\\
%\end{tabular}
\hline\\
\multicolumn{5}{c}{10 Fold CV}\\\\
\hline
Efficiency Measure &Probit&Logit&Cauchit&C-Log-Log\\
\hline
$E_1$&0.89928&0.89928&0.89208&0.90647\\
$E_2$&0.78801&0.85911&0.82494&0.78705\\
$E_3$&0.94224&0.93713&0.87470&0.93064\\
$E_4$&0.88832&0.89685&0.84048&0.83493\\
\hline
\end{longtable}
\begin{table}[ht]
\caption{Description and classification of the covariates}
%\resizebox{}{}{}
\resizebox{!}{0.65\textwidth}{\begin{tabular}{lll}
\hline
\\
{\textbf{Covariates}} & {\textbf{Description}} & \textbf{Type} \\ 
\\
\hline
\\
GDP per Capita & \begin{tabular}[c]{@{}l@{}}Purchasing Power Parity as given by World Development Indicators\end{tabular}  & Strength \\
\\
\hline
\\
Social support & \begin{tabular}[c]{@{}l@{}}It is the national average of the binary responses (either 0 \\ or 1) to the GWP question “If you were in trouble, do\\  you have relatives or friends you can count on to help\\  you whenever you need them, or not?”\end{tabular}  & Strength \\ 
\\
\hline
\\
Healthy Life Expectancy & \begin{tabular}[c]{@{}l@{}}The time series of healthy life expectancy at birth\\ are based on data  from the World Health Organization\\  (WHO),  the World Development Indicators (WDI), \\ and statistics published in journal articles  taken as\\  non-health adjusted life expectancy and adjusted the\\  time series of total life expectancy to healthy life \\ expectancy by simple multiplication, assuming that \\ the ratio remains constant within each country over the\\ sample period.\end{tabular}& Strength \\ 
\\
\hline
\\
Freedom to make life choices & \begin{tabular}[c]{@{}l@{}}It is the national average of responses to the GWP question “Are \\ you satisfied or dissatisfied with your freedom to choose what\\ you do with your life?”\end{tabular}  & Strength \\ \\
\hline
\\
Generosity & \begin{tabular}[c]{@{}l@{}}It is the residual of regressing national average of response to the \\ GWP question “Have you donated money to a charity in the past \\ month?” on GDP per capita.\end{tabular}  & Strength \\ 
\\
\hline
\\
Corruption Perception & \begin{tabular}[c]{@{}l@{}}The measure is the national average of the survey responses \\ to two questions in the GWP: “Is corruption widespread\\ throughout the government or not” and “Is corruption \\ widespread within businesses or not?” The overall perception\\  is just the average of the two 0-or-1 responses.\end{tabular} & Stress \\ 
\\
\hline
\\
Positive affect & \begin{tabular}[c]{@{}l@{}}It is defined as the average of three positive affect measures in\\ GWP: happiness, laugh and enjoyment in the Gallup World Poll waves 3-7.\end{tabular} & Strength \\ 
\\
\hline
\\
Negative affect & \begin{tabular}[c]{@{}l@{}}It is defined as the average of three negative affect measures in\\ GWP, worry, sadness and anger,\end{tabular} & Stress \\ 
\\
\hline
\\
Confidence in national government & \begin{tabular}[c]{@{}l@{}}GWP asked the question that “Do you have confidence in\\ each of the following, or not? How about the national government?\end{tabular} & Strength \\
\\
\hline
\\
Democratic and Delivery Quality & \begin{tabular}[c]{@{}l@{}}This is based on WGI, which accounts Voice and Accountability, Political\\ Stability and Absence of Violence, Government Effectiveness, Regulatory\\ Quality, Rule of Law, Control of Corruption. The indicators are on a scale\\ roughly with mean zero and a standard deviation of 1. In WHR to reduce the \\ dimensions to two using the simple average of the first two measures as an \\ indicator of democratic quality, and the simple average of the other four measures\\ as an indicator of delivery quality.\end{tabular} & Strength \\ 
\\
\hline
\end{tabular}}
\end{table}

\end{document}